\documentclass[showpacs,amsmath,amssymb,aps,preprint]{revtex4}
\usepackage{graphicx}
\usepackage{bm}
\usepackage[utf8]{inputenc}
\usepackage{lmodern}
\usepackage{graphicx}
\begin{document}
\title{Relativistic Klein-Gordon-Maxwell multistream model for quantum plasmas}

\author{F. Haas}
\affiliation{Departamento de F{\'i}sica, Universidade Federal do Paran\'a, 81531-990, Curitiba, Paran\'a, Brazil}
\author{B. Eliasson}
\affiliation{International Centre for Advanced Studies in Physical Sciences and Institute for Theoretical
Physics, Faculty of Physics \& Astronomy, Ruhr University Bochum, D-44780 Bochum, Germany}
\author{P. K. Shukla}
\affiliation{International Centre for Advanced Studies in Physical Sciences and Institute for Theoretical
Physics, Faculty of Physics \& Astronomy, Ruhr University Bochum, D-44780 Bochum, Germany, and
Department of Mechanical and Aerospace Engineering \& Center for Energy Research, University of California
San Diego, La Jolla, CA 92093, U. S. A.}
\received{18 February 2012}
\begin{abstract}
A multistream model for spinless electrons in a relativistic quantum plasma is introduced by means of a suitable fluid-like version of the Klein-Gordon-Maxwell system. The one and two-stream cases are treated 
in detail. A new linear instability condition for two-stream quantum plasmas is obtained, generalizing the previously known non-relativistic results. In both the one and two-stream cases, steady-state solutions reduce the model to a set of coupled nonlinear ordinary differential equations, which can be numerically solved, yielding a manifold of nonlinear periodic and soliton structures. The validity conditions for the applicability of the model are addressed.
\end{abstract}
\pacs{52.27.Ny, 52.35.Qz, 52.35.Sb}
\maketitle
\section{Introduction}
The interest in relativistic quantum plasma systems is growing exponentially, not only because of the relevance to astrophysical problems but also due to the fast advances in strong laser-solid plasma interaction experiments. Indeed, the development of multi-Peta-Watt lasers will soon make possible to address simultaneous quantum and relativistic effects in laboratory plasmas \cite{Mourou}. Relativistic quantum kinetic models have been proposed, in the treatment of the plasma dispersion function for a Fermi-Dirac equilibrium \cite{Melrose1}, and in the study of relativistic effects for quantum ion-acoustic wave propagation \cite{Melrose2}. Also, a covariant Wigner function theory for relativistic quantum plasmas described by the Dirac-Maxwell system has been suggested \cite{Hakim}, as well as the spinless (Klein-Gordon-Maxwell) analog has been presented \cite{Mendonca}. The Klein-Gordon-Maxwell system of equations has been applied to the analysis of parametric scattering instabilities in relativistic laser-quantum plasma interactions \cite{Bengt1}, while recent models have been introduced based on the Dirac-Maxwell equations describing the nonlinear propagation of light in Dirac matter \cite{Bengt2}.

Relativistic two-stream instabilities are traditionally known to be important for the electron heating in 
intense laser-plasma interaction experiments \cite{Thode73}, as well as in astrophysical relativistic shocks \cite{Nakar11}, and could be important for pulsar glitches \cite{Anderson03,Anderson04}, where superfluid neutrons and superconducting protons co-exist with relativistic electrons \cite{Samuelson10}. In addition, the two-stream instabilities between electrons and/or holes are also believed to exist in semiconductor plasmas \cite{Stenflo68,Robinson67,Grib00}.

In a lowest level of approximation than kinetic theory, quantum plasma hydrodynamic models are popular tools (see,  e.g. \cite{Shukla10, Shukla11, Vladimirov11, Haas11}), since they allow an efficient treatment of nonlinear phenomena, both from the a\-na\-ly\-ti\-cal and numerical viewpoints. Starting from the Dirac-Maxwell system, there are hydrodynamic models for relativistic quantum plasmas \cite{Taka55}, which have been extended to incorporate particle-antiparticle effects in the wave propagation \cite{Asenjo}, as well as relativistic quantum corrections to laser wakefield acceleration \cite{Zhou}. In these fluid formulations, we note, in particular, the non-trivial form of the re\-la\-ti\-vis\-tic extension of the 
quantum tunneling force in the momentum transport equation. Electromagnetic quantum hydrodynamic wave equations have also been considered in\-clu\-ding relativistically degenerate electron fluids \cite{Masood}, based on previous spin-one-half hydrodynamic  models \cite{Brodin07,Marklund07,Shukla09,Lundin10}. 
We note that hydrodynamic versions of the Klein-Gordon-Maxwell system have been used in the past \cite{Taka53},   and recently in the context of laser physics \cite{Andreev}.

 In this paper, we introduce a relativistic multistream quantum plasma model starting from the Klein-Gordon-Maxwell system of equations.  We adapt the formalism according to the classical \cite{Dawson} and quantum \cite{Haas00} multistream model for plasmas, and show its usefulness in a paradigmatic plasma problem, namely the linear and non\-li\-near features of the quantum two-stream instability in the relativistic regime.  The simplicity of the Klein-Gordon equations, in comparison to  the Dirac equation, makes it a natural candidate for the extension of non-relativistic quantum plasma theories to the relativistic regime, when electron-one-half spin effects can be neglected. Therefore, a direct comparison to existing results on quantum plasmas can be obtained in an easier way. Moreover, an hydrodynamic formulation based on the Klein-Gordon-Maxwell system of equations strongly favors the development of new analytical and numerical tools for relativistic quantum plasmas. On the other hand, the domain of applicability is restricted to plasmas where the electron-one-half spin effects are not decisive, like in close to isotropic equilibrium configurations. For the parameters in this work, we have no quantized electromagnetic fields, so that quantum field theoretic results involving e.g. pair creation are not included. Streaming instabilities in non-relativistic quantum plasmas are attracting considerable interest, since they display many surprising characteristics of pure quantum origin. Among these, we have a new instability branch for large wavenumbers, as well as new nonlinear spatially periodic solutions in the steady state case \cite{EPL}. Besides formulating the relativistic version of the quantum multistream model for plasmas, the purpose of the present work is to extend the analysis of the quantum two-stream instability to the relativistic case.

The  manuscript is organized as follows. In Section II we introduce the multistream Klein-Gordon-Maxwell system of equations, casting it into a suitable fluid-like formulation. The one-stream case is treated in detail in Section III, where linear wave propagation is studied considering small amplitude perturbations around homogeneous equilibria, and a rich variety of nonlinear periodic as well as soliton structures are found numerically. An existence criterion for solitary wave solutions is obtained. In Section IV, the relativistic quantum two-stream case is studied in depth. The linear relativistic quantum two-stream instability problem is fully characterized. In this regard, a main result of this work is the derivation of the instability condition 
in Eq. (\ref{insta}), which provides a natural generalization to the non-relativistic instability criterion \cite{Haas00}. Both nonlinear periodic and soliton structures are found numerically, and an existence criterion for localized solutions is obtained theoretically. Section V is dedicated to the final conclusions, including a detailed account on the validity domain of our model equations.

\section{The Klein-Gordon-Maxwell multistream model}

We here consider a relativistic multi-stream quantum plasma where the electrons are described by a statistical mixture of $N$ pure states, with each wavefunction $\psi_j$ satisfying the Klein-Gordon 
equation

\begin{equation}
\label{e1}
{\cal W}^2 \psi_j - c^2 {\cal P}^2 \psi_j - m^2 c^4 \psi_j = 0\,, \quad j=1,\, \ldots,N,
\end{equation}
where we have defined the energy and momentum operators respectively as

\begin{equation}
{\cal W} = i\hbar\frac{\partial}{\partial t} + e\phi,
\end{equation}
and
\begin{equation}
{\cal P} = - i\hbar\nabla + e{\bf A} \,.
\end{equation}
Here, $m$ and $-e$ are the electron mass and charge, respectively, $\hbar$ is the Planck constant divided by $2\pi$, $c$  is the speed of light in vacuum, and $\phi$ and ${\bf A}$ are the scalar and vector potentials, respectively.

The electric charge and current densities are, respectively,

\begin{equation}
\rho = - \frac{e}{2mc^2}\sum_{j=1}^N\left[\psi_j^{*}{\cal W}\psi_j + \psi_j({\cal W}\psi_j)^{*}\right],
\end{equation}
and

\begin{equation}
{\bf J}  = - \frac{e}{2m}\sum_{j=1}^N \left[\psi_j^{*}{\cal P}\psi_j + \psi_j({\cal P}\psi_j)^{*}\right] \,.
\end{equation}

The charge and current densities fulfill the continuity equation

\begin{equation}
\frac{\partial\rho}{\partial t} + \nabla\cdot{\bf J} = 0 \,.
\end{equation}

The self-consistent scalar and vector potentials are obtained from the inhomogeneous Maxwell's equations,
using the Coulomb gauge $\nabla \cdot {\bf A}=0$, as

\begin{eqnarray}
\label{p}
\nabla^2\phi = - \frac{1}{\varepsilon_0}(\rho + n_0 e) \,,\\
\label{a}
\Box{\bf A} = \mu_0 {\bf J} - \frac{1}{c^2}\nabla\frac{\partial\phi}{\partial t} \,,
\end{eqnarray}
where a fixed neutralizing ion background of the charge density $e n_0$ was added, and
where $\varepsilon_0$ and $\mu_0$ denote the vacuum electric permittivity and magnetic permeability, respectively. Here, the d'Alembert operator is 

\begin{equation}
\Box = \frac{1}{c^2}\frac{\partial^2}{\partial t^2} - \nabla^2 \,.
\end{equation}

The resulting Klein-Gordon-Maxwell system of equations (\ref{e1}) and (\ref{p})--(\ref{a}) describes 
the nonlinear interactions in relativistic quantum plasmas where spin effects  and pair creation phenomena are negligible. In explicit form, the Klein-Gordon equation reads

\begin{equation}
\Box\psi_j - \frac{ie}{\hbar c^2}\bigg(\frac{\partial\phi}{\partial t}\psi_j + 2\phi\frac{\partial\psi_j}{\partial t} + 2c^2{\bf A}\cdot\nabla\psi_j\bigg) + \frac{1}{\hbar^2}\bigg(e^2 A^2 - \frac{e^2\phi^2}{c^2}+m^2 c^2\bigg)\psi_j = 0 \,.
\end{equation}

Following Takabayasi \cite{Taka53}, it is convenient to introduce a fluid-like formulation in terms of the eikonal decomposition

\begin{equation}
\psi_j = R_j \exp(iS_j /\hbar),
\end{equation}
where the amplitude $R_j$ and phase $S_j$ are real functions. Separating the real and imaginary parts of the Klein-Gordon equation, we have 

\begin{equation}
\frac{1}{c^2}\bigg(\frac{\partial S_j}{\partial t} - e\phi\bigg)^2 - (\nabla S_j + e{\bf A})^2 - m^2 c^2 = \frac{\hbar^2 \Box R_j}{R_j},
\end{equation}
and

\begin{equation}
R_j\bigg(\Box S_j - \frac{e}{c^2}\frac{\partial\phi}{\partial t}\bigg) + \frac{2}{c^2}\frac{\partial R_j}{\partial t}\bigg(\frac{\partial S_j}{\partial t} - e\phi\bigg) - 2\nabla R_j\cdot(\nabla S_j + e{\bf A}) = 0 \,.
\end{equation}

In terms of $R_j$ and $S_j$, the charge and current densities are, respectively,

\begin{equation}
\rho = \frac{e}{mc^2} \sum_{j=1}^N R_j^2 \bigg(\frac{\partial S_j}{\partial t} - e\phi\bigg),
\end{equation}
and

\begin{equation}
{\bf J} = - \frac{e}{m} \sum_{j=1}^N R_j^2 (\nabla S_j + e{\bf A}) \,.
\end{equation}

Alternative hydrodynamic-like methods are also available, as for instance the Feshbach-Villars formalism \cite{Feshbach58} where initially the Klein-Gordon equation is split into a pair of first-order in time partial differential equations. However, the resulting set of equations turns out to appear much more involved than in the present Takabayasi approach, which we use due to its formal simplicity.

From now on we concentrate on the electrostatic case, and assume ${\bf A} = 0$. In this situation, the relevant equations are

\begin{eqnarray}
\label{hj_m}
\frac{1}{c^2}\bigg(\frac{\partial S_j}{\partial t} - e\phi\bigg)^2 - (\nabla S_j)^2 - m^2 c^2 = \frac{\hbar^2 \Box R_j}{R_j} \,,
\\
R_j \bigg(\Box S_j - \frac{e}{c^2}\frac{\partial\phi}{\partial t}\bigg) + \frac{2}{c^2}\frac{\partial R_j}{\partial t}\bigg(\frac{\partial S_j}{\partial t} - e\phi\bigg) - 2\nabla R_j \cdot\nabla S_j = 0,
\end{eqnarray}
and

\begin{equation}
\label{ps_m}
\nabla^2\phi = - \frac{e}{\varepsilon_0}\left[\sum_{j=1}^N\frac{R_j^2}{mc^2}\bigg(\frac{\partial S_j}{\partial t} - e\phi\bigg) + n_0\right] \,.
\end{equation}
The assumption ${\bf A}=0$ is valid as long as the electric current is curl free, $\nabla\times {\bf J}=0$. 
This is true for longitudinal waves in a stationary plasma and parallel to a plasma beam.

\section{One-stream case}

We proceed next to study linear and nonlinear waves for the one-stream case ($N=1$).
For this case the sum in Eq. (\ref{ps_m}) collapses to one term involving $R_1=R$
and $S_1=S$, and Eqs. (\ref{hj_m})--(\ref{ps_m}) become

\begin{eqnarray}
\label{hj}
\frac{1}{c^2}\bigg(\frac{\partial S}{\partial t} - e\phi\bigg)^2 - (\nabla S)^2 - m^2 c^2 
= \frac{\hbar^2 \Box R}{R} \,,
\\
R\bigg(\Box S - \frac{e}{c^2}\frac{\partial\phi}{\partial t}\bigg) + \frac{2}{c^2}\frac{\partial R}{\partial t}\bigg(\frac{\partial S}{\partial t} - e\phi\bigg) - 2\nabla R\cdot\nabla S = 0,
\end{eqnarray}
and

\begin{equation}
\label{ps}
\nabla^2\phi = - \frac{e}{\varepsilon_0}\left[\frac{R^2}{mc^2}\bigg(\frac{\partial S}{\partial t} - e\phi\bigg) + n_0\right] \,.
\end{equation}

\subsection{Linear waves}

The system of equations (\ref{hj})--(\ref{ps}) has the equilibrium solution

\begin{equation}
R = \sqrt{\frac{n_0}{\gamma}} \,, \quad S = - \gamma mc^2 t + {\bf p}\cdot{\bf r} \,, \quad \phi = 0 \,,
\end{equation}
where

\begin{equation}
\gamma = \left(1+\frac{p^2}{m^2 c^2}\right)^{1/2}
\end{equation}
is the relativistic $\gamma$ factor for a beam momentum ${\bf p}$. Linearizing and assuming perturbations $\sim \exp(i[{\bf K}\cdot{\bf r} - \Omega t])$, where, for simplicity, we take ${\bf K}\parallel{\bf p}$, 
and obtain

\begin{equation}
\label{serb}
(\Omega - Kv)^2 = \frac{\omega_{p}^2}{\gamma^3} + \frac{\hbar^2}{4\gamma^2 m^2}\left(K^2-\frac{\Omega^2}{c^2}\right)^2 + \frac{\hbar^2 \omega_{p}^2}{4 \gamma^3 m^2 c^2}\left(K^2 -\frac{\Omega^2}{c^2}\right) \,,
\end{equation}
where $v = p/(\gamma m)$ is the equilibrium beam speed and $\omega_p = (n_0 e^2/(m\varepsilon_0))^{1/2}$.

In the non-streaming limit $p\rightarrow 0$ (and $\gamma=1$), Eq.~(\ref{serb}) is identical to
Eq.~(4.21) of Kowalenko {\it et al.} \cite{Kowalenko85}.

For the general one-stream case with waves propagating obliquely to the beam direction,
the assumption ${\bf A}=0$ fails, and one has to involve the full set of Maxwell's equations.
However, in the one-stream case, one can start with the 3D dispersion
relation in the beam frame, and then Lorentz transform the result to the
laboratory frame. In the beam frame, the dispersion relation is

\begin{equation}
(\Omega')^2=   (\omega_{p}')^2
+\frac{\hbar^2}{4 m^2} \bigg[(K')^2-\frac{(\Omega')^2}{c^2}\bigg]^2
+\frac{\hbar^2(\omega_{p}')^2}{4 m^2 c^2} \bigg[(K')^2-\frac{(\Omega')^2}{c^2}\bigg],
\end{equation}
where the primed $\Omega'$ and ${\bf K}'$ are the angular frequency and wave vector of the plasma oscillations in the beam frame.

To go from the beam frame to the la\-bo\-ra\-to\-ry frame, we assume for simplicity that the beam velocity is along the $z$ axis. Then, the time and space variables are Lorentz transformed
as $t'=\gamma (t-v z/c^2)$, $x'=x$, $y'=y$ and $z'=\gamma(z-v t)$.
The corres\-pon\-ding frequency and wavenumber transformations are $\Omega'=\gamma(\Omega-v K_z)$, $K_x'=K_x$, $K_y'=K_y$, and $K_z'=\gamma(K_z-{v \Omega}/{c^2})$.

The plasma frequency is transformed as $\omega_{p}'=\omega_{p}/\sqrt{\gamma}$.
One easily verifies that the expression $(K')^2-(\Omega')^2/c^2=K^2-\Omega^2/c^2$
is Lorentz invariant. This yields imme\-dia\-te\-ly the general dispersion relation for beam oscillations 
in the laboratory frame,

\begin{equation}
  (\Omega-v K_z)^2=
\frac{\omega_{p}^2}{\gamma^3}+\frac{\hbar^2}{4 \gamma^2 m^2} \bigg(K^2-\frac{\Omega^2}{c^2}\bigg)^2
+\frac{\hbar^2 \omega_p^2}{4 \gamma^3 m^2 c^2} \bigg(K^2-\frac{\Omega^2}{c^2}\bigg).
\end{equation}

In the formal classical limit ($\hbar = 0$), we have from Eq.~(\ref{serb}) the Doppler shifted relativistic
plasma oscillations \cite{Godfrey75,Yan86} $\Omega = Kv + \omega_{p}\gamma^{-3/2}$.
Using the limit $\Omega \approx Kv$ and $\Omega \gg \omega_p$ in the right-hand side of 
Eq.~(\ref{serb}) one obtains

\begin{equation}
(\Omega - Kv)^2 = \frac{\omega_{p}^2}{\gamma^3} + \frac{\hbar^2 K^4}{4\gamma^6 m^2} \,,
\end{equation}
which is similar to the expression used by Serbeto {\it et al.}~\cite{Serbeto09} in the context of quantum
free-electron lasers, and where the last term in the right-hand side can be considered a quantum correction
to the relativistic beam-plasma mode.

On the other hand, in the non-relativistic limit $c \rightarrow \infty$ we have $\gamma = 1$ and the 
familiar result \cite{Haas00}

\begin{equation}
(\Omega - K v)^2 = \omega_{p}^2 + \frac{\hbar^2 K^4}{4 m^2} \,,
\end{equation}
describing Doppler-shifted quantum Langmuir waves.

Normalizing into dimensionless units according to

\begin{equation}
\Omega^* = \frac{\Omega}{\omega_p} \,, \quad K^* = \frac{cK}{\omega_p} \,, \quad p^* = \frac{p}{mc} \,, \quad v^* = \frac{v}{c} \,, \quad  H = \frac{\hbar\omega_{p}}{mc^2},
\end{equation}
one obtains (omitting the asterisks)

\begin{equation}
\label{dis}
(\Omega - Kv)^2 = \frac{1}{\gamma^3} + \frac{H^2}{4\gamma^2}(K^2-\Omega^2)^2 + \frac{H^2}{4\gamma^3}(K^2-\Omega^2) \,,
\end{equation}
where now $\gamma = (1+p^2)^{1/2} = (1-v^2)^{-1/2}$.
If $H \neq 0$ and  $v = 0$, we solve Eq. (\ref{dis}) to find the modes $\Omega = \Omega_{\pm}$, with

\begin{equation}
\Omega_{\pm}^2 = \frac{1}{2}+K^2+\frac{2}{H^2} \pm 2\left[\frac{K^2}{H^2}+\left(\frac{1}{4}-\frac{1}{H^2}\right)^2\right]^{1/2} \,.
\end{equation}
It can be verified that both modes are stable ($\Omega^{2}_{\pm} > 0$). It is reasonable to expand the last result assuming a small $H$. Even for laser-compressed matter in the laboratory \cite{Azechi91,Kodama01},
the value of $H$ presently does not significantly exceed $10^{-3}$.  On the other hand, for conditions in the interior 
of white dwarf stars with a quantum coupling parameter exceeding unity we reach the pair creation regime \cite{Tsytovich61}, which can be safely treated only within the quantum field theory. The result is

\begin{eqnarray}
\Omega_{+}^2 = \frac{4}{H^2} + 2K^2 + \frac{H^2 K^2}{4}(1-K^2) + O(H^4) \,,\\
\Omega_{-}^2 = 1 - \frac{H^2 K^2}{4}(1-K^2) + O(H^4) \,.
\end{eqnarray}
Here $\Omega_+$ is the pair branch which goes to infinity as $H \rightarrow 0$.

In dimensional units, the pair branch has a cutoff at $\Omega= 2 m c^2/\hbar$.
The $\Omega_-$ is actually a backward wave (negative group velocity) for small wavenumbers,
as mentioned by Kowalenko {\it et al.}~\cite{Kowalenko85} below their Eq.~(4.23).

\subsection{Nonlinear stationary solutions}

Next, we consider nonlinear stationary solutions of Eqs.~(\ref{hj})--(\ref{ps}) in one spatial dimension, 
of the form

\begin{equation}
R = R(x) \,, \quad S = - \gamma mc^2 t + S_{0}(x) \,, \quad \phi = \phi(x) \,,
\end{equation}
so that the original partial differential equation system is converted into a system of ordinary differential equations

\begin{eqnarray}
\label{rrr}
\hbar^2 c^2 R'' = \left[(S_{0}')^2 c^2 - p^2 c^2 - 2 e \gamma m c^2 \phi - e^2\phi^2\right] R \,,\\
\label{con}
R S_{0}'' + 2 R' S_{0}' = 0 \,, \\
\label{ff}
\phi'' = \frac{e}{\varepsilon_0} (\gamma R^2 - n_0) + \frac{\omega_{p}^2}{n_0 c^2} R^2 \phi \,,
\end{eqnarray}
where the primes denote derivatives with respect to $x$.

Equation (\ref{con}) can be immediately integrated as $R^2 S_0'=$constant.
This relation is also equivalent to current continuity $J_x=$constant, which
follows from the continuity equation (6) with $\partial \rho / \partial t=0$. Assuming that $R=\sqrt{n_{0}/\gamma}$ and $S_{0}'=p$ where the plasma is at equilibrium, we have

\begin{equation}
R^2 S_{0}' = \frac{n_0 p}{\gamma} \Rightarrow S_{0}' = \frac{n_0 p}{\gamma R^2} \,,
\label{S0prime}
\end{equation}
which inserted into Eq.~(\ref{rrr}) yields

\begin{equation}
\label{RR}
\hbar^2 c^2 R'' + \left(p^2 c^2 + 2 e \gamma m c^2 \phi + e^2\phi^2\right) R = \frac{n_0^2 p^2 c^2}{\gamma^2 R^3} \,.
\end{equation}

Equations (\ref{ff}) and (\ref{RR}) form a coupled nonlinear system for $\phi$ and $R$,
describing steady state solutions of our relativistic quantum plasma.

Other special solutions (traveling wave solutions, alternative boundary conditions) could also be investigated, but we keep the above scheme, since then we can  directly compare to the previous linear wave analysis. Indeed, Eqs.~(\ref{ff}) and (\ref{RR}) admit the equilibrium

\begin{equation}
\label{eq}
R^2 = \frac{n_0}{\gamma} \,, \quad \phi = 0 \,,
\end{equation}
in the same way as the original Klein-Gordon-Maxwell system of equations (which also needs the equilibrium phase $S$).

To proceed, we first transform into dimensionless variables according to

\begin{equation}
R^* = \frac{R}{\sqrt{n_0}} \,, \quad \phi^* = \gamma + \frac{e\phi}{mc^2} \,, \quad x^* = \frac{\omega_p x}{c} \,, \quad p^* =  \frac{p}{mc}\,, \quad
S_0^*=\frac{\omega_p S_0}{m c^2} \,,
\end{equation}
so that the system for stationary waves becomes (omitting the asterisks)

\begin{eqnarray}
\label{Rbis}
H^2 R'' + (\phi^2 - 1) R = \frac{v^2}{R^3} \,, \\
\label{phibis}
\phi'' = R^2 \phi - 1 \,,
\end{eqnarray}
and

\begin{equation}
  S_0'=\frac{v}{R^2},
\end{equation}
where $v = p/\gamma$.

In the system (\ref{Rbis}) and (\ref{phibis}), the variable $x$ takes the role of a time-like variable, so that standard methods for ordinary differential equations can be applied. In this context, it is interesting to investigate the system around the equilibrium point if it admits only oscillatory (stable) solutions, or if it also admits exponentially growing (unstable) and decaying solutions. In the latter case, there is a possibility of finding localized solitary waves solutions with exponentially decaying flanks, which is not possible for stable cases.

Linearizing the system around the equilibrium (\ref{eq}) and supposing perturbations $\propto \exp(iKx)$, we obtain the characteristic equation for the eigenvalues $K$

\begin{equation}
\frac{1}{\gamma}\bigg(\frac{H^2 K^2}{4}-p^2\bigg)\bigg(K^2 + \frac{1}{\gamma}\bigg) + 1 = 0 \,,
\end{equation}
which is the same as Eq.~(\ref{serb}) with $\Omega = 0$. In the formal classical limit ($H = 0$), we have only linearly stable oscillations with

\begin{equation}
\label{kv}
K^2 v^2 = \frac{1}{\gamma^3} \,.
\end{equation}

Hence, in the classical case, as is well-known, we do not have localized stationary solutions.
In the quantum case ($H \neq 0$) the situation is more complex. The cha\-rac\-te\-ris\-tic equation
 can be solved yielding $K^2 = K_{\pm}^2$, with

\begin{equation}
K_{\pm}^2 = \frac{2}{\gamma H^2}\bigg\{\gamma p^2 - \frac{H^2}{4} \pm \bigg[\bigg(\frac{H^2}{4}+\gamma p^2\bigg)^2-\gamma^3 H^2\bigg]^{1/2}\bigg\} \,.
\label{Kpm}
\end{equation}

Expanding for small $H$, we obtain

\begin{eqnarray}
K_{+}^2 = \frac{4p^2}{H^2} - \frac{\gamma}{p^2}-\frac{H^2}{4p^6} + O(H^4) \,,\\
K_{-}^2 = \frac{1}{\gamma p^2} + \frac{H^2}{4p^6} + O(H^4) \,.
\end{eqnarray}
Note that $K_{-}$ is the quantum extension of the branch in Eq.~(\ref{kv}), while $K_+$ has no classical analog.

To investigate the stability of the equilibrium, it is useful to rewrite the characteristic equation as

\begin{equation}
F(K^2) = \frac{\gamma}{(p^2-H^2 K^2/4)(K^2+1/\gamma)} = 1 \,,
\end{equation}
which gives a second degree equation for $K^2$. Since the characteristic function $F(K^2)$ satisfies

\begin{equation}
F(0) = \frac{\gamma^2}{p^2} \geq 1 \,, \quad F\bigg(K^2 > \frac{4p^2}{H^2}\bigg) < 0 \,,
\end{equation}
and has a pole at $K^2 = 4p^2/H^2$,

\begin{equation}
\lim_{K^2 \rightarrow (4p^2/H^2)^{\mp}} F(K^2) = \pm \infty \,,
\end{equation}
we will have stable oscillations provided that the minimum value $F_{\rm min}$ in the branch 
$0 \leq K^2 < 4p^2/H^2$ satisfies

\begin{equation}
\label{min}
F_{\rm min} < 1 \,.
\end{equation}
In this case $F(K^2)$ intercept the value $1$ at two positive $K^2$ values, so that the characteristic 
equation has only real solutions. The situation is summarized in Fig.~1.

\begin{figure}[!ht]
\centering{\includegraphics[width=10.0cm]{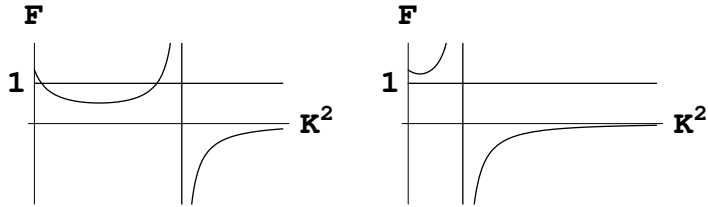}
\caption{on the left, a typical case of stable linear one-stream oscillations when $F_{\rm min} < 1$. 
On the right, a typical unstable case where $F_{\rm min} > 1$. The pole is at $K^2 = 4p^2/H^2$}
\label{fig1}}
\end{figure}

Working out Eq.~(\ref{min}) we find that

\begin{equation}
\gamma H^2 < \bigg(p^2 + \frac{H^2}{4\gamma}\bigg)^2 \,. 
\label{stability}
\end{equation}

\begin{figure}[!ht]
\centering{\includegraphics[width=10.0cm]{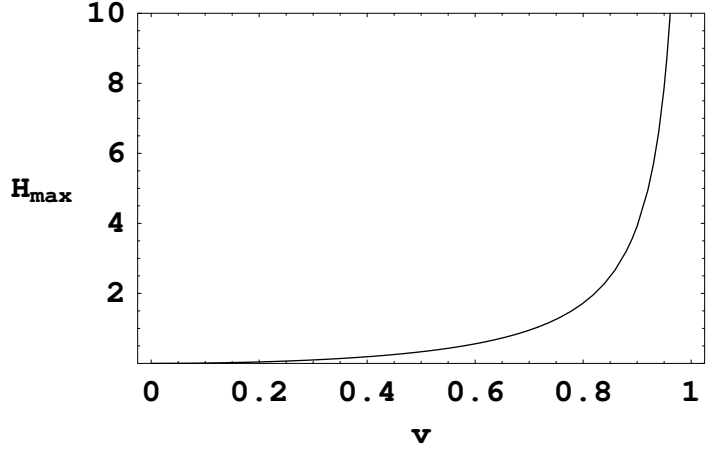}
\caption{maximum quantum parameter $H_{\rm max}$ in Eq.~(\ref{h}) as a function of the velocity $v$ measured in units of $c$}
\label{fig2}}
\end{figure}

Further analysis shows that the condition (\ref{stability}) can be written as

\begin{equation}
\label{h}
H < H_{\rm max} = 2\gamma (\gamma^{1/2}-\gamma^{-1/2}) \,, 
\end{equation}
as the final condition for stable linear oscillations. It follows that the necessary existence criterion for
nonlinear localized (soliton) stationary solutions is $H > H_{\rm max}$. In Fig.~\ref{fig2}, a graph of
$H_{\rm max}$ is plotted as a function of the velocity $v$ (measured in units of $c$), which shows
that the quantum range for stable oscillations is increased for increasing relativistic effects.
In the non-relativistic limit $p\ll 1$, we have $H < p^2=v^2$ as the condition for stable oscillations, 
or in dimensional units, $\hbar \omega_p< m v^2$.

For the nonlinear system (\ref{Rbis}) and (\ref{phibis}), a Hamiltonian form can be obtained with 
the further transformation

\begin{equation}
R \rightarrow iR \,, \quad \phi \rightarrow \frac{\phi}{H} \,, \quad x \rightarrow x \,,
\end{equation}
so that

\begin{equation}
R'' = - \frac{\partial V}{\partial R} \,, \quad \phi'' = - \frac{\partial V}{\partial\phi} \,,
\end{equation}
where

\begin{equation}
V = V(R,\phi) = \frac{R^2\phi^2}{2} + \frac{\phi}{H} - \frac{R^2}{2H^2} + \frac{v^2}{2H^2 R^2} \,.
\end{equation}

Since we arrive at an autonomous Hamiltonian system, one has the energy integral

\begin{equation}
I = \frac{(\phi')^2}{2} + \frac{(R')^2}{2} + V(R,\phi) \,.
\end{equation}

Restoring dimensional variables, we have the conserved quantity

\begin{equation}
\tilde{I} = \frac{2\hbar^2\omega_{p}^2}{m^2 c^4}I - 2\gamma \,,
\end{equation}
or

\begin{equation}
\begin{split}
\tilde{I} &= \frac{\hbar^2}{m^2 c^2} \left[ \left(\frac{e}{m c^2}\frac{d\phi}{dx}\right)^2 - \frac{1}{n_0}\bigg(\frac{dR}{dx}\bigg)^2 \right]
\\
&+ \frac{2e\phi}{mc^2} - \frac{R^2}{n_0}\bigg(\gamma + \frac{e\phi}{mc^2}\bigg)^2 
+ \frac{R^2}{n_0} - \frac{n_0 \beta^2}{R^2} \,,
\end{split}
\end{equation}
where $\beta = v/c$. The obtained conservation law can be used to verify the accuracy of numerical simulations.

\begin{figure}[htb]
\centering
\includegraphics[width=10cm]{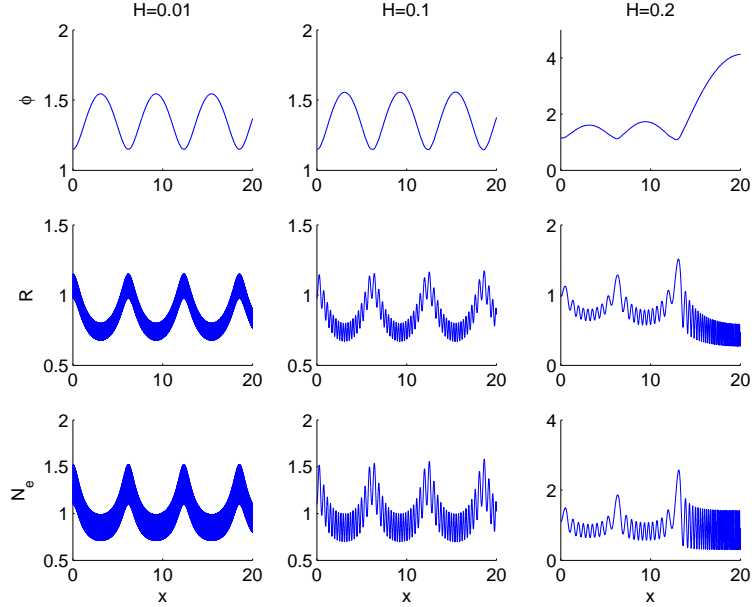}
\caption{Spatial profiles of $\phi$, $R$, and the electron number density $N_e=R^2\phi$ (top to bottom panels) for $\gamma=1.3$, and $H=0.01$ (left column), $H=0.1$ (middle column) and $H=0.2$ (right column). The solution was set to $\phi(0)=1.15$, $R(0)=0.98$ and $\phi'(0)=R'(0)=0$ at the left boundary}
\label{onestream_sim}
\end{figure}

\begin{figure}[htb]
\centering
\includegraphics[width=10cm]{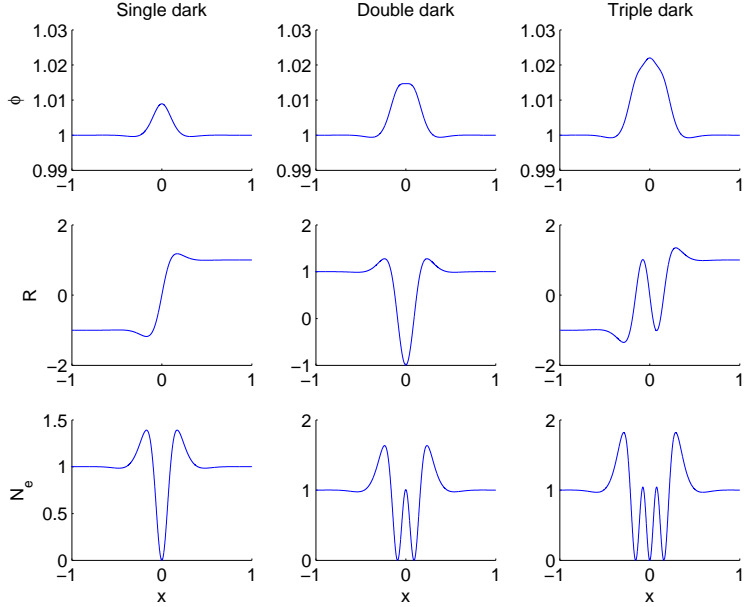}
\caption{Spatial profiles of $\phi$, $R$, and the electron density $N_e=R^2\phi$ (top to bottom panels), showing single, double and triple dark solitary waves (left to right columns) for the zero beam speed case $v=0$ with $H=0.01$. The solution is set to $\phi=|R|=1$ at the left and right boundaries
($R=-1$ on the left boundary for the single and triple dark solitons)}
\label{fig_H_0_01}
\end{figure}

\begin{figure}[htb]
\centering
\includegraphics[width=10cm]{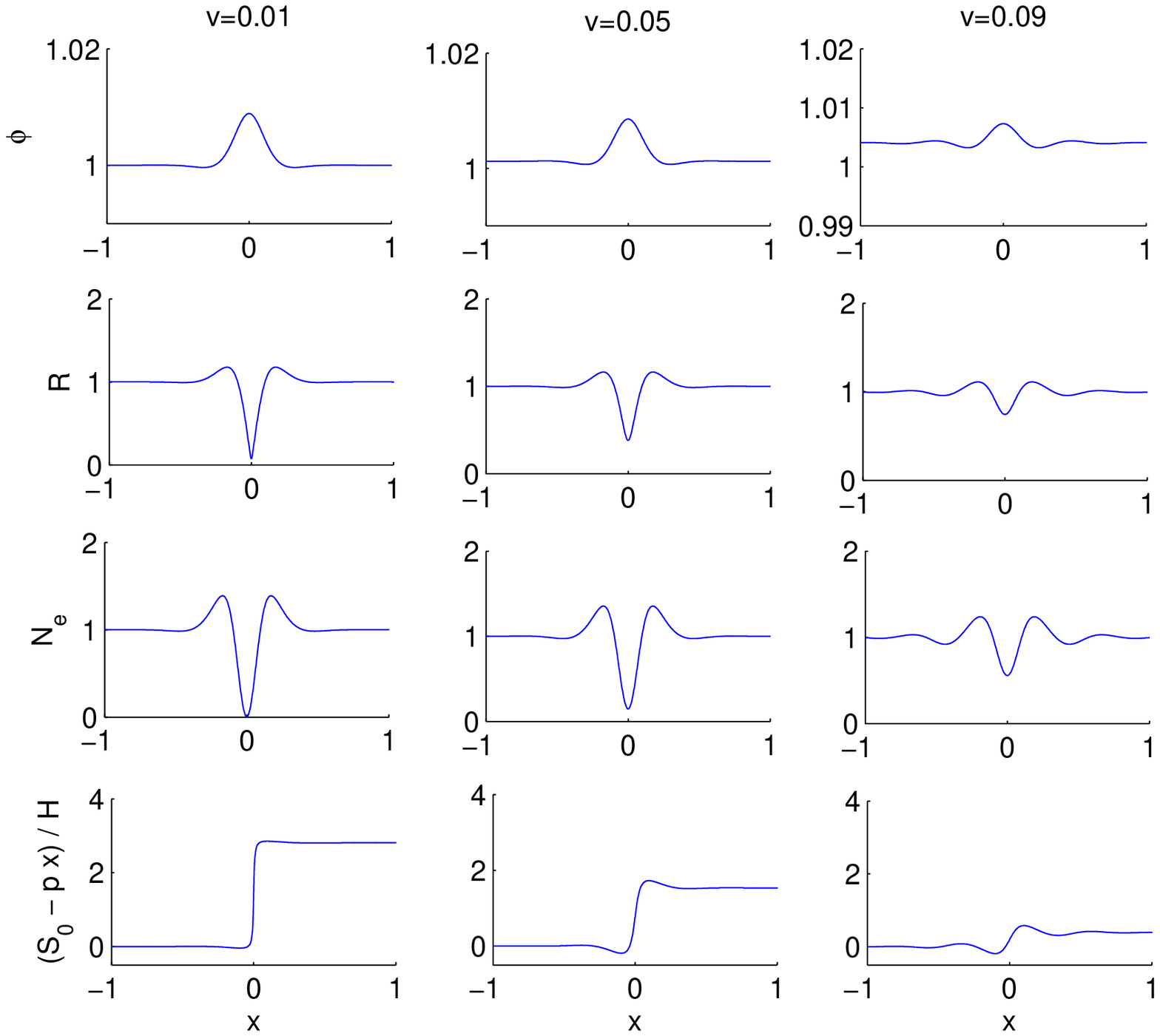}
\caption{spatial profiles of $\phi$, $R$, $N_e=R^2\phi$, and $(S_0-px)/H$ (top to bottom panels), for
$H=0.01$, and $v=0.01$ (left column), $v=0.05$ (middle column),
and $v=0.09$ (right column). The solution was set to $\phi=\gamma$
and $R=1/\sqrt{\gamma}$ at the left and right boundaries. We see grey solitons
with non-zero electron density in the center. Bottom panels show the phase shift $(S_0 - p x)/H$ 
of the wavefunction}
\label{fig_H_0_01_v0}
\end{figure}

\begin{figure}[htb]
\centering
\includegraphics[width=10cm]{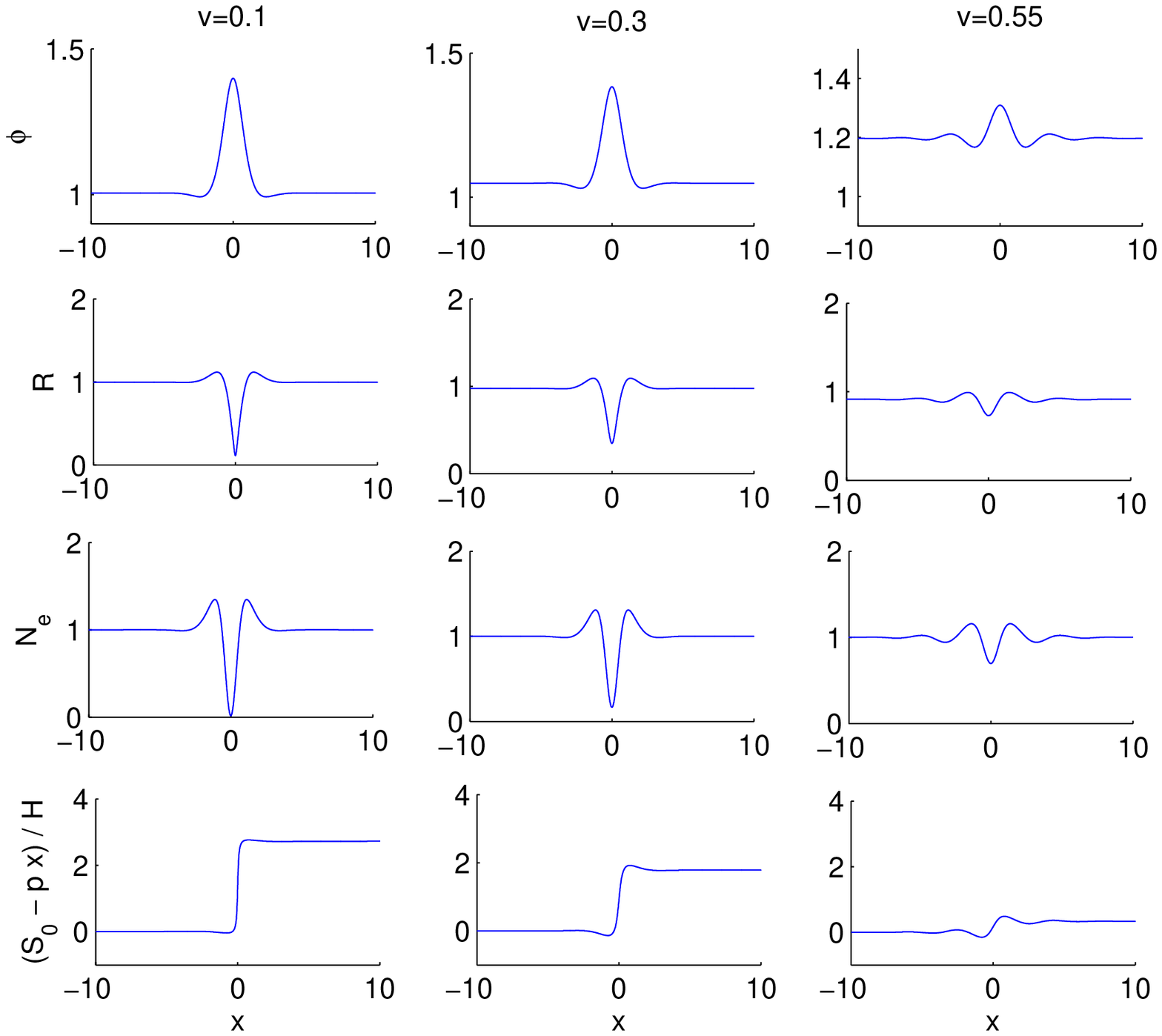}
\caption{Spatial profiles of $\phi$, $R$, $N_e=R^2\phi$, and $(S_0-px)/H$ (top to bottom panels), for
$H=0.5$, and $v=0.1$ (left column), $v=0.3$ (middle column),
and $v=0.55$ (right column). The solution was set to $\phi=\gamma$
and $R=1/\sqrt{\gamma}$ at the left and right boundaries. We see grey solitons
with non-zero electron density in the center. Bottom panels show the phase shift $(S_0 - p x)/H$ of the wavefunction}
\label{fig_H_0_5_v0}
\end{figure}

Numerical solutions of the nonlinear system (\ref{Rbis}) and (\ref{phibis}) are presented
in Figs.~\ref{onestream_sim}--\ref{fig_H_0_5_v0}. For the non-localized solutions
in Fig.~\ref{onestream_sim}, initial values on $\phi$, $R$, and their first derivatives were
set on the left boundary and solution was integrated using the standard 4th-order Runge-Kutta
method. For the localized solutions in Figs. \ref{fig_H_0_01}--\ref{fig_H_0_5_v0}, boundary
conditions on $\phi$ and $R$ were fixed on both the left and right boundaries, and the solutions
were found with iterations based on Newton's method. Figure \ref{onestream_sim} shows large
amplitude oscillations for $\gamma=1.3$ and different values of $H$, such that
small-amplitude oscillations are linearly stable, in the sense discussed above. We see
that there is one short and one long length-scale, corresponding to $K_+$ and
$K_-$ for the linear oscillations in Eq.~(\ref{Kpm}). Here the small-scale
oscillations are due to the quantum diffraction effect, while the large-scale
oscillations are related to wakefield oscillations which are well-known
in classical plasmas \cite{Berezhiani99}. When these two length-scale
become comparable, i.e. for large $H$, the two length-scales interact
and the oscillations become more irregular.

For parameters where the oscillations are exponentially decaying or increasing (unstable oscillations),
we have the possibility of localized solutions in the form of dark or grey solitons.
It turns out that the coupled system of equations (\ref{Rbis}) and (\ref{phibis}) supports a
wide variety of nonlinear localized structures. Due to quantum diffraction effects,
the plasma can develop dark solitary waves with one or more electron density minima.
In Fig.~\ref{fig_H_0_01}, we see different classes of dark solitary waves for the case when the plasma is at rest, $v=0$, and $H=0.01$.  Since $p=v=0$, in this case, the term proportional to $1/R^3$ in the right-hand side of Eq.~(\ref{Rbis}) vanishes, and $R$ can continuously go between positive and negative values.
We see single, double and triple dark solitons, where $R$ is shifted 180 degrees  (from negative to positive) on between the left and right sides of the single and triple dark solitons. The dark soliton with single electron minimum is the same type as found in Ref. \cite{Shukla06} for a non-relativistic quantum plasma. The solutions with multiple density minima are somewhat similar in shape  to the multiple-hump optical solitons predicted in relativistic laser-plasma interactions in the classical regime \cite{Kaw92,Saxena06}. On the other hand, in Figs.~\ref{fig_H_0_01_v0} and \ref{fig_H_0_5_v0}, we consider solitons in a streaming plasma with finite speed $v > 0$. We recall that the existence condition for solitons is
$H > 2\gamma (\gamma^{1/2}-\gamma^{-1/2})$ [where $\gamma=1/(1-v^2)^{1/2}$], which puts an upper limit on $v$ for a given value of $H$. For example, for $H=0.01$, shown in Fig. \ref{fig_H_0_01_v0}, we have
$v \lesssim 0.1$, while for $H=0.5$, shown in Fig.  \ref{fig_H_0_5_v0},
we have $v \lesssim 0.58$ for solitons to exist. For $H \ll 1$ and non-relativistic $v\ll 1$, the existence condition for solitary structures becomes $v^2<H$, or in dimensional units, $m v^2 < \hbar \omega_p$.
A general feature of the propagating solitons is that the electron density is non-zero at the center of the soliton, hence they are grey solitons. Furthermore, as the speed increases, the amplitudes of the solitons decrease and their tails become oscillatory when the speed approaches the maximum
allowed speed, as can be seen in the right-hand columns of Figs.~\ref{fig_H_0_01_v0} and \ref{fig_H_0_5_v0}. There is also a complex phase shift proportional to $S_0$ in the total wave function $\psi$ due to the relation (\ref{S0prime}). The plot of $(S_0 - p x)/H$ in Fig.~\ref{fig_H_0_5_v0} shows how the phase (in radians) is shifted between the two sides of the solitons. As $v\rightarrow 0$, the phase jumps abruptly 
a value of $\approx \pi$ in the center of the soliton, while solitons with higher speeds have smaller and smoother jumps in the phase, as can be seen in the bottom panels of Figs.~\ref{fig_H_0_01_v0} and \ref{fig_H_0_5_v0}.

\section{The two-stream case}

We next consider linear and nonlinear waves for the two-stream case ($N=2$). For this case, stream is represented by a wavefunction $\psi_j = R_j\exp(iS_j/\hbar), \, j = 1,2$, and we have the Klein-Gordon-Poisson system of equations

\begin{eqnarray}
\label{r1}
\frac{1}{c^2}\left(\frac{\partial S_j}{\partial t} - e\phi\right)^2 - (\nabla S_j)^2 - m^2 c^2 = \frac{\hbar^2 \Box R_j}{R_j} \,, \\
\label{s1}
R_j \left(\Box S_j - \frac{e}{c^2}\frac{\partial\phi}{\partial t}\right) + \frac{2}{c^2}\frac{\partial R_j}{\partial t}\left(\frac{\partial S_j}{\partial t}-e\phi\right) - 2\nabla R_j\cdot\nabla S_j  = 0 \,,\\
\label{f1}
\nabla^2 \phi = - \frac{e}{\varepsilon_0}\left[\frac{1}{mc^2}\sum_{j=1}^{2} R_{j}^2\left(\frac{\partial S_j}{\partial t} - e\phi\right) + n_0 \right] \,,
\end{eqnarray}
which describe a relativistic quantum two-stream plasma in the electrostatic approximation, using physical variables.

\subsection{Linear waves}

Similar to the one-stream case, we have the equilibrium

\begin{eqnarray}
R_1 = R_2 = \left(\frac{n_0}{2\gamma}\right)^{1/2} , \quad \phi = 0 \,, \quad \gamma = \left(1+\frac{p^2}{m^2 c^2}\right)^{1/2} \,, \nonumber \\
\quad S_1 = - \gamma mc^2 t + {\bf p}\cdot{\bf r} \,, \quad S_2 = - \gamma mc^2 t - {\bf p}\cdot{\bf r} \,,
\end{eqnarray}
for two symmetric counter-propagating electron streams.

Linearizing the governing equations and assuming plane-wave perturbations with the wavenumber 
${\bf K}\parallel{\bf p}$ and the angular frequency $\Omega$, we obtain  the dispersion relation

\begin{equation}
\label{dre}
F(\Omega) = 1  \,,
\end{equation}
with the characteristic function $F(\Omega)$ defined by

\begin{equation}
\label{cf}
F(\Omega) =  \frac{\omega_{b}^2}{\gamma} \, \sum_{+,-} \frac{4m^2 c^4 - \hbar^2 (\Omega^2-c^2 K^2)}{4\gamma^2 m^2 c^4 (\Omega \mp Kv)^2 - \hbar^2 (\Omega^2 - c^2 K^2)^2} \,,
\end{equation}
where

\begin{equation}
\omega_b = \left(\frac{n_0 e^2}{2m\varepsilon_0}\right)^{1/2} \,, \quad v = p/(\gamma m) \,.
\end{equation}

In the classical limit, viz. $\hbar = 0$,  we obtain the same results as in the description of classical cold relativistic electron beams using the fluid theory \cite{Thode73}. It can be treated in full analytical detail. 
We have the dispersion relation

\begin{equation}
\label{cl}
\Omega^2 = K^2 v^2 + \frac{\omega_{b}^2}{\gamma^3} \pm \frac{\omega_b}{\gamma^3} (\omega_{b}^2 + 4\gamma^3 K^2 v^2)^{1/2} \,.
\end{equation}

\begin{figure}[!ht]
\centering{\includegraphics[width=10cm]{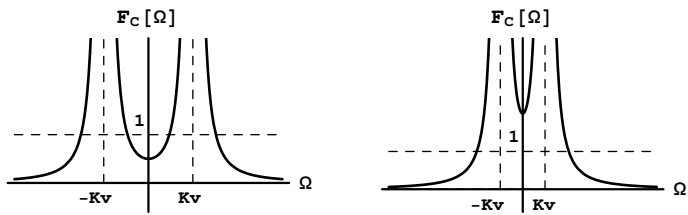}
\caption{On the left, stable linear waves satisfying $F_{C}(0) < 1$, with the non-quantum characteristic function $F_{C}(\Omega)$ given by Eq.~(\ref{fc}). On the right, unstable linear waves}
\label{fig3}}
\end{figure}

Equivalently, it is useful to write the classical dispersion relation as $F_{C}(\Omega) = 1$, with the characteristic function

\begin{equation}
\label{fc}
F_{C}(\Omega) = \frac{\omega_{b}^2}{\gamma^3}\left[\frac{1}{(\Omega - Kv)^2} + \frac{1}{(\Omega + Kv)^2}\right] \,,
\end{equation}
obtained by setting $\hbar = 0$ in Eq.~(\ref{cf}). The dispersion relation turns out to be a quadratic equation for $\Omega^2$, hence $F_{C}(\Omega)$ should attain the unity value four times to prevent instability. Graphically (see Fig.~\ref{fig3}) we conclude that

\begin{equation}
F_{C}(0) < 1
\end{equation}
is the condition for linear stability. This shows that the wave-numbers such that

\begin{equation}
K^2 v^2 > \frac{\omega_{p}^2}{\gamma^3}
\end{equation}
are linearly stable. The same conclusion is reached analyzing the potentially unstable mode in Eq.~(\ref{cl}). In comparison with the non-relativistic stability condition $K^2 > \omega_{p}^2/v^2$, we note that the relativistic effects are stabilizing, since they imply a smaller unstable range in wave-number space.

Setting $\Omega = i\Omega_i$ for real $\Omega_i$ and using Eq.~(\ref{cl}), we obtain

\begin{equation}
{\rm max}(\Omega_i) = \frac{\omega_b}{2\gamma^{3/2}}
\end{equation}
as the maximum growth rate, which also becomes smaller due to relativistic effects.

On the other hand, in the quantum but non-relativistic ($\hbar \neq 0$, and $c \rightarrow \infty$) limit
 we have

\begin{equation}
1 - \sum_{+,-}\frac{\omega_{b}^2}{(\Omega \mp Kv)^2 - \hbar^2 K^4/(4m^2)} = 0 \,.
\end{equation}

We will not discuss the non-relativistic case, since this has been already done in the past \cite{Haas00}. 
The non-relativistic case, as well as the non-quantum case, can be solved in full analytical detail, because 
in both situations the dispersion relation is equivalent to a second degree polynomial equation for $\Omega^2$.

We now turn our attention to the fully quantum-relativistic dispersion relation (\ref{dre}). Due to the symmetry, we can restrict the treatment to positive frequencies, wave-numbers and beam velocities. Equation (\ref{dre}) is equivalent to a fourth degree polynomial equation for $\Omega^2$, which can be analytically solved in terms of cumbersome expressions, or solved numerically. However, it is more informative to first analyze the behavior of the characteristic function $F(\Omega)$ in Eq.~(\ref{cf}), which is mainly determined by the poles at

\begin{eqnarray}
\Omega_1 &=& \frac{\gamma mc^2}{\hbar} - \left[\left(\frac{\gamma mc^2}{\hbar}-Kv\right)^2+\frac{K^2 c^2}{\gamma^2}\right]^{1/2} \,,\\
\Omega_2 &=& - \frac{\gamma mc^2}{\hbar} + \left[\left(\frac{\gamma mc^2}{\hbar}+Kv\right)^2+\frac{K^2 c^2}{\gamma^2}\right]^{1/2} \,,\\
\Omega_3 &=& \frac{\gamma mc^2}{\hbar} + \left[\left(\frac{\gamma mc^2}{\hbar}-Kv\right)^2+\frac{K^2 c^2}{\gamma^2}\right]^{1/2} \,,\\
\Omega_4 &=& \frac{\gamma mc^2}{\hbar} + \left[\left(\frac{\gamma mc^2}{\hbar}+Kv\right)^2+\frac{K^2 c^2}{\gamma^2}\right]^{1/2} \,,
\end{eqnarray}
paying attention just to the positive values. More precisely, $\Omega_1 > 0$ provided $\hbar K < 2\,p$, otherwise the positive pole is at $-\Omega_1$. It can be shown that one has the ordering

\begin{equation}
|\Omega_1| < \Omega_2 < \Omega_3 < \Omega_4 \,.
\end{equation}
Here, $\Omega_1$ and $\Omega_2$ have classical counterparts as $\hbar\rightarrow 0$, while $\Omega_3$ and $\Omega_4$ are associated with pair branches, without classical counterparts.

We note that the case $\hbar K = 2 p$ is degenerate, since then one has $\Omega_1 = 0$ and the dispersion relation becomes a third degree polynomial equation for $\Omega^2$. The solutions to this particular case always correspond to (marginally) stable modes, not considered any further here.

A tedious analysis shows that the characteristic function has the following properties

\begin{eqnarray}
\lim_{\Omega \rightarrow \Omega_{1}^{\pm}} F(\Omega) = \mp \infty \,, \quad \lim_{\Omega \rightarrow \Omega_{2, 3, 4}^{\pm}} F(\Omega) = \pm \infty \,, \nonumber \\
\hbar K < 2p \Rightarrow F\left(\Omega_1 < \Omega < \Omega_2\right) < 0 \\
{\rm Sign} F(0) = - {\rm Sign} F''(0) = - {\rm Sign}(\hbar K - 2p) \,. \nonumber
\end{eqnarray}

\begin{figure}[!ht]
\centering{\includegraphics[width=10cm]{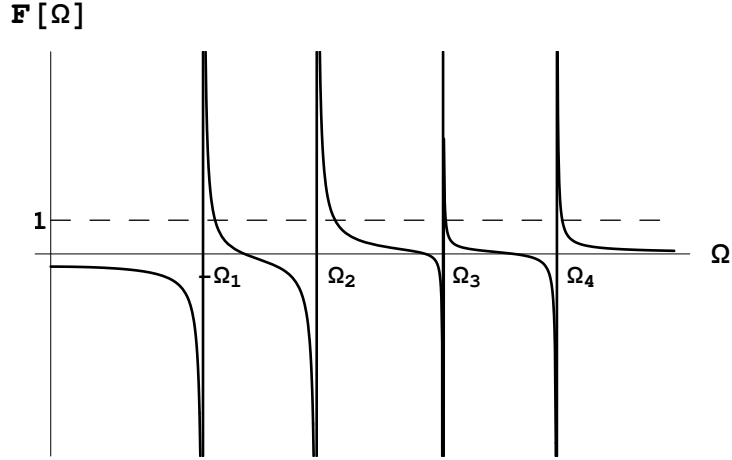}
\caption{Generic behavior of the characteristic function $F(\Omega)$ in Eq.~(\ref{cf}) for $\hbar K > 2 p$. Since $F = 1$ at four positive frequencies, the corresponding wavenumber is stable}
\label{fig4}}
\end{figure}

Moreover, $F(\Omega)$ tend monotonously to zero as $\Omega \rightarrow \infty$. These results imply that the wave-numbers satisfying $\hbar K > 2 p$ are always stable, since in this case the characteristic function has the topology shown in Fig.~\ref{fig4}, where $F(\Omega)$ always intercepts the value unity four times.

\begin{figure}[!ht]
\centering{\includegraphics[width=10cm]{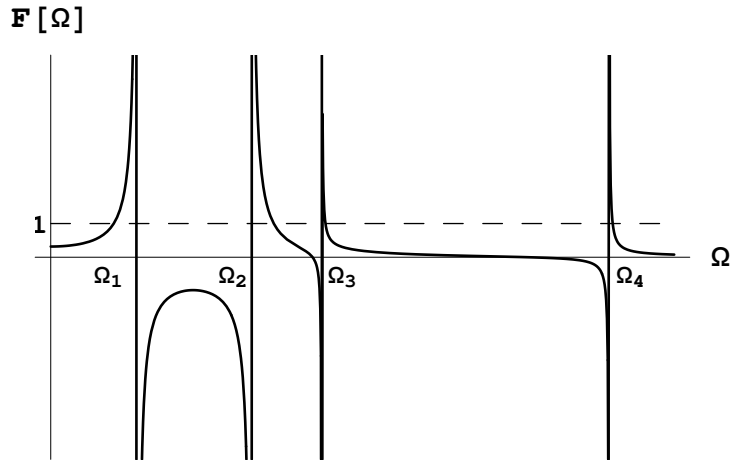}
\caption{Generic behavior of the characteristic function $F(\Omega)$ in Eq.~(\ref{cf}) for $\hbar K < 2 p$, in the stable cases where $F(0) < 1$}
\label{fig5}}
\end{figure}

\begin{figure}[!ht]

\centering{\includegraphics[width=10cm]{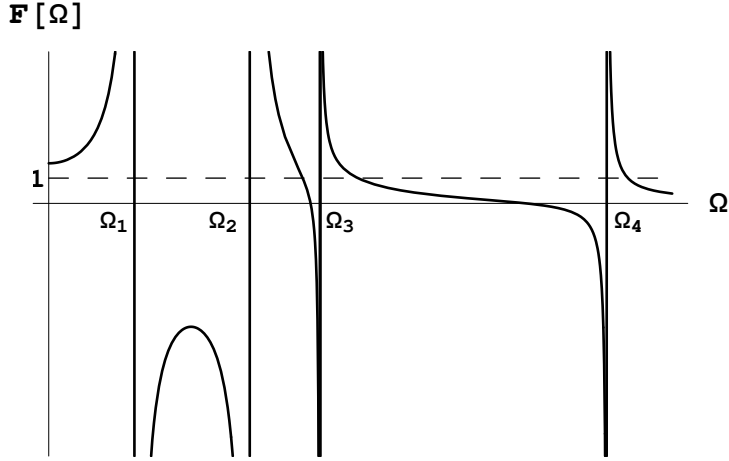}
\caption{Generic behavior of the characteristic function $F(\Omega)$ in Eq.~(\ref{cf}) for $\hbar K < 2 p$, in the unstable cases where $F(0) > 1$}
\label{fig6}}
\end{figure}

On the other hand, the case $\hbar K < 2p$ is potentially unstable, according to the minimum value $F(0)$. If $F(0) < 1$, the characteristic function attains the unity value at four positive frequencies, corresponding to four linearly stable waves. Otherwise, when $F(0) > 1$ there is a (purely imaginary) solution for the dispersion relation, and hence instability. The whole scenario is summarized in Figs.~\ref{fig5} and \ref{fig6}.


In summary, besides $\hbar K < 2p$ we have

\begin{equation}
F(0) = \frac{\omega_{p}^2}{\gamma^3 K^2 v^2} \,\, \frac{1 + \hbar^2 K^2/(4 m^2 c^2)}
{1 - \hbar^2 K^2/(4 p^2)} > 1,
\end{equation}
as a necessary condition for unstable linear wave propagation in our two-stream relativistic quantum plasma. Rearranging the instability conditions found, we combine them according to

\begin{equation}
\label{insta}
4p^2 > \hbar^2 K^2 > 4\,\left(\frac{\gamma K^2 v^2/\omega_{p}^2-1+\beta^2}{\gamma K^2 v^2/\omega_{p}^2+\beta^2}\right)\,p^2 \,.
\end{equation}

It can be verified that Eq.~(\ref{insta}) reproduces the non-relativistic results \cite{Haas00}.

\begin{figure}[!ht]
\centering{\includegraphics[width=12cm]{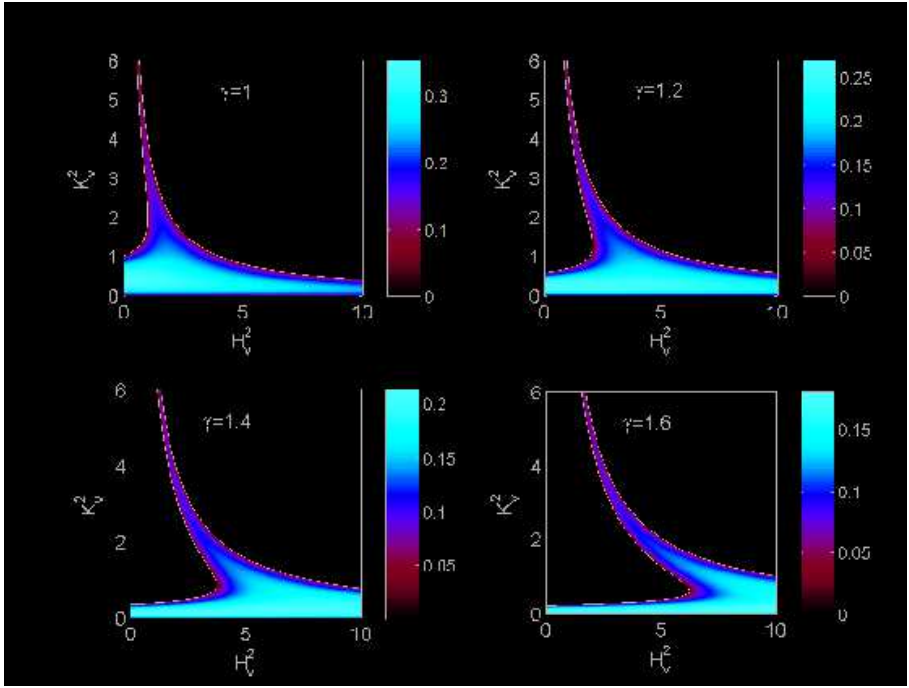}
\caption{Unstable cases as a function of $H_v^2=\hbar^2\omega_p^2/m^2 v^4$ and $K_v^2=K^2 v^2/\omega_p^2$ for different values of $\gamma$, where the unstable cases, given by Eq. (\ref{insta}) are between the lower and upper boundaries of the colored region. The color corresponds to the normalized growth rate $\Omega_i/\omega_p$ obtained numerically from Eq.~(\ref{dre}).
The case $\gamma=1$ (upper left panel) is the non-relativistic case corresponding to Fig.~1 of Ref. \cite{Haas00}. For increasing values of $\gamma$, the instability region is shifted towards larger values of $H_v^2$ and the growth rate decreases}
\label{fig7}}
\end{figure}

The instability condition (\ref{insta}), together with a numerical solution of
Eq.~(\ref{dre}) are depicted in Fig.~\ref{fig7}. Here we assumed that $\Omega=\Omega_r + i\Omega_i$, where $\Omega_r$ is the real frequency and $\Omega_i$ the growth rate, and plotted $\Omega_i/\omega_p$ as a function of $H_v^2=\hbar^2\omega_p^2/m^2 v^4$ and $K_v^2=K^2 v^2/\omega_p^2$ for different values of $\gamma$. The case $\gamma=1$ corresponds to Fig.~1 of Ref. \cite{Haas00}, while $\gamma>1$ show the relativistic effects on the instability region. We note that in the formal classical limit the largest unstable wavenumber becomes smaller as $\gamma \rightarrow \infty$. On the other hand, the height of the upper curve in the instability diagram scales as $\gamma^2$, so that in this sense the combined quantum-relativistic effects tend to enlarge the unstable area.
Ultimately, however, quantum effects stabilize the sufficiently small wave numbers, no matter the strength of relativistic effects. An interesting quantum effect is the appearance of an instability region at large wavenumbers $K_v^2=K^2 v^2/\omega_p^2$ for moderately small values of $H_v^2=\hbar^2\omega_p^2/m^2 v^4$, which does not have a classical counterpart. Finally, it should be noted that simultaneously $H_v^2\gtrsim 1$ and $\gamma > 1$ in Fig.~\ref{fig7} cor\-res\-pond to extremely high electron number densities, comparable to those in the interiors of white dwarf stars and similar astrophysical objects.

\subsection{Nonlinear stationary solutions}

We consider the one-dimensional version of the system (\ref{r1})--(\ref{f1}) and stationary solutions of the form

\begin{equation}
R_{1,2} = R_{1,2}(x) \,, \quad S_{1,2} = - \gamma mc^2 t + \sigma_{1,2}(x) \,, \quad \phi = \phi(x) \,.
\end{equation}

Equations (\ref{s1}) are then equivalent to

\begin{equation}
\label{fin}
\frac{d}{dx}(R_{1}^2 \sigma_{1}') = \frac{d}{dx}(R_{2}^2 \sigma_{2}') = 0 \,,
\end{equation}
where the primes denote $x-$derivatives. Assuming

\begin{equation}
R_{1}^2 = R_{2}^2 = \frac{n_0}{2\gamma} \,, \quad \sigma_{1}' = - \sigma_{2}' = p
\end{equation}
at equilibrium, applying the transformation

\begin{equation}
R_{1,2}^{*} = \frac{R_{1,2}}{\sqrt{n_0}} \,, \quad \phi^* = \gamma + \frac{e\phi}{mc^2} \,, \quad x^* = \frac{\omega_b x}{c},
\end{equation}
and using Eq.~(\ref{fin}) to eliminate $\sigma_{1,2}'$ from Eq.~(\ref{r1}), we readily derive the system of equations (omitting the asterisks)

\begin{eqnarray}
\label{R1bis_ts}
H^2 R_{1}'' + (\phi^2 -1) R_1 = \frac{v^2}{4R_{1}^3} \,,\\
H^2 R_{2}'' + (\phi^2 -1) R_2 = \frac{v^2}{4R_{2}^3} \,,
\end{eqnarray}
and

\begin{eqnarray}
\label{phibis_ts}
\phi'' = (R_{1}^2 + R_{2}^2)\,\phi - 1,
\end{eqnarray}
which predict nonlinear stationary solutions of a relativistic quantum two-stream plasma. Here$H$ and $v$ are defined as in the one-stream case.

\begin{figure}[htb]
\centering
\includegraphics[width=10cm]{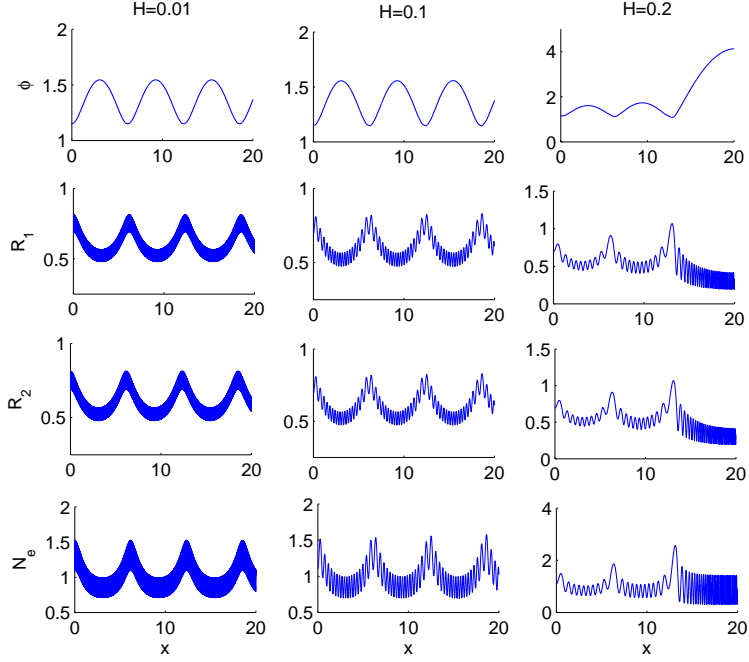}
\caption{Spatial profiles of $\phi$, $R_1$, $R_2$, and the electron density $N_e=(R_1^2+R_2^2)\phi$ (top to bottom panels) for $H=0.01$ (left column), $H=0.1$ (middle column) and $H=0.2$ (right column).
The solution was set to $\phi(0)=1.15$, $R_1(0)=R_2(0)=0.98/\sqrt{2}$ and $\phi'(0)=R_1'(0)=R_2'(0)=0$ at the left boundary}
\label{twostream_sim1}
\end{figure}

\begin{figure}[htb]
\centering
\includegraphics[width=10cm]{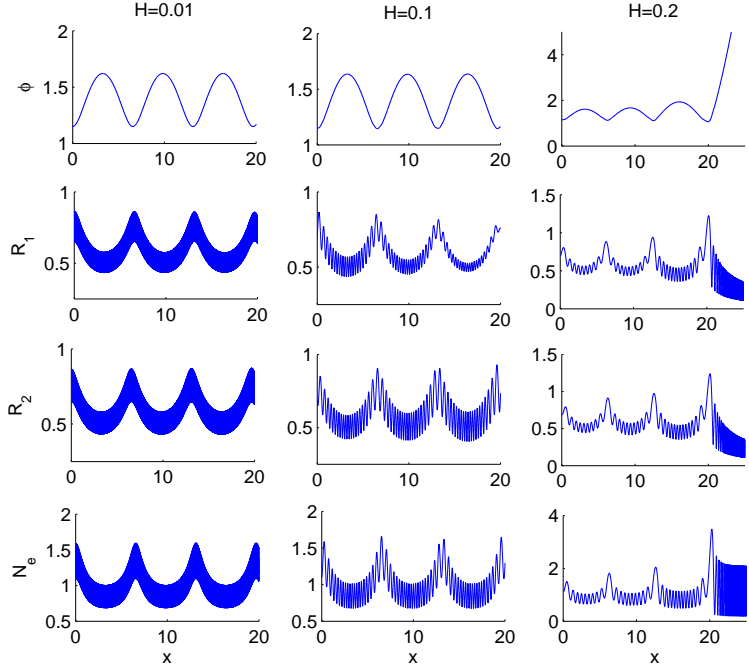}
\caption{The same as in Fig. \ref{twostream_sim1}, but using $R_1'(0)=10$ and
$R_2'(0)=-10$ (left column), $R_1'(0)=1$ and $R_2'(0)=-1$ (middle column), and $R_1'(0)=0.1$ and
$R_2'(0)=-0.1$ (right column)}
\label{twostream_sim2}
\end{figure}

Linearizing around $R_{1,2}^2 = 1/(2\gamma)$ and $\phi = \gamma$, and supposing perturbations $\propto \exp(iKx)$, we obtain a quadratic equation for $K^2$, which is also obtained from Eq.~(\ref{dre}) setting $\Omega = 0$. Proceeding as before, we  formally obtain the same existence condition as Eq. (\ref{h})
for periodic solutions. The two-stream nonlinear solutions can be constructed from the one-stream cases by using $|R_1|=|R_2|=|R|/\sqrt{2}$, where $R$ is obtained by solving the system (\ref{Rbis})--(\ref{phibis}). The signs of $R_1$ and $R_2$ are arbitrary. In Fig. \ref{twostream_sim1}, we show a numerical solution of the system (\ref{r1})--(\ref{phibis_ts}), where the profiles of $\phi$ and $N_e$ are identical to the ones in Fig. \ref{onestream_sim}, and with $R_1=R_2=R/\sqrt{2}$. In Fig. \ref{twostream_sim2}, we perturbed this solution by using different values of $R_1'(0)$ and $R_2'(0)$ at the left boundary. The general behavior of the solution in Fig. \ref{twostream_sim2} remains similar as in Fig. \ref{twostream_sim1}, but differences in the two solutions can be seen in the details.

\begin{figure}[htb]
\centering
\includegraphics[width=10cm]{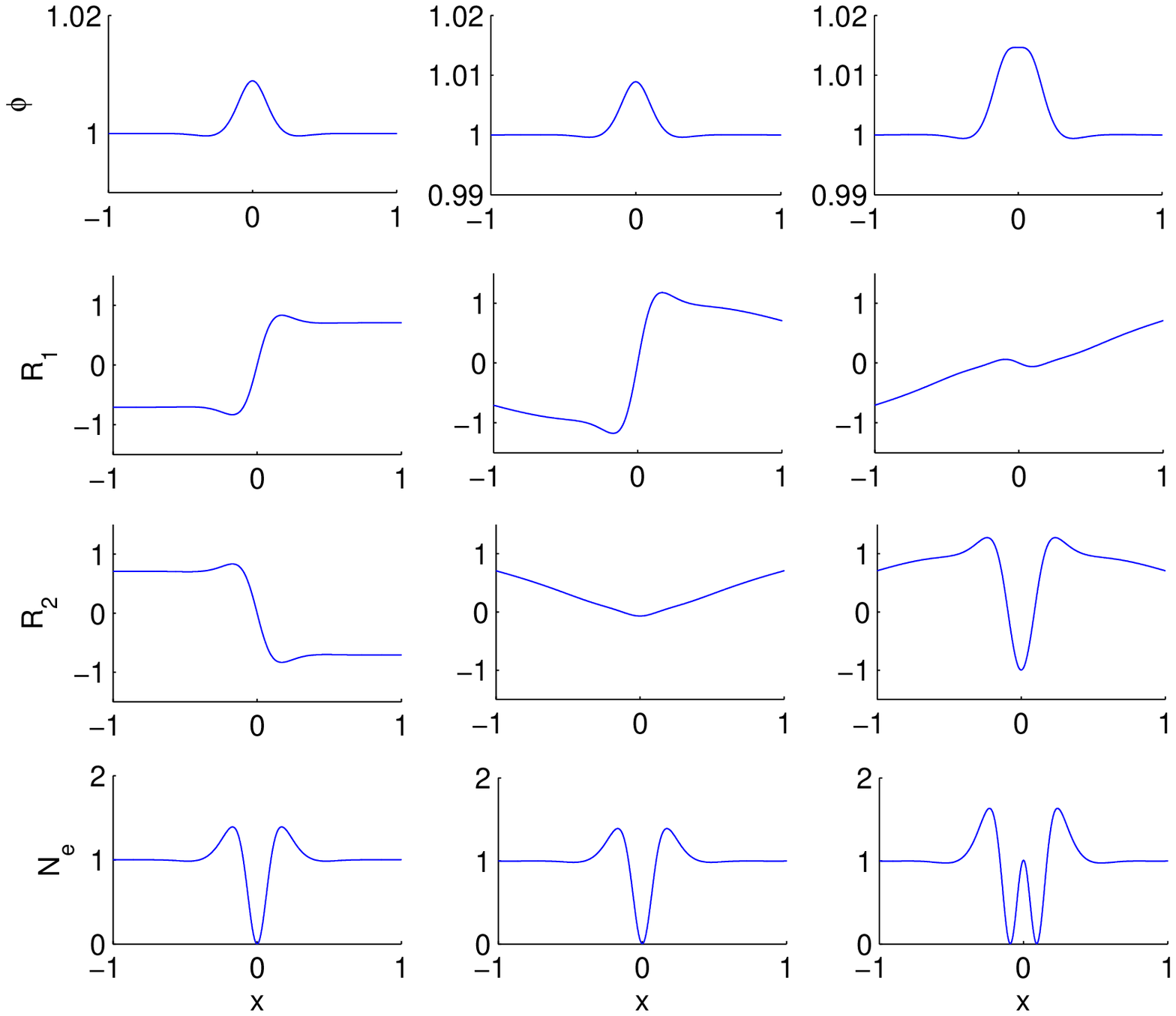}
\caption{Spatial profiles of $\phi$, $R_1$, $R_2$, and $N_e=(R_1^2+R_2^2)\phi$ (top to bottom panels)
for the zero beam speed case $v=0$ with $H=0.01$. The solution is set to $\phi=|R_1|=|R_2|=1/\sqrt{2\gamma}$ at the left and right boundaries}
\label{twostream_stationary}
\end{figure}

\begin{figure}[htb]
\centering
\includegraphics[width=10cm]{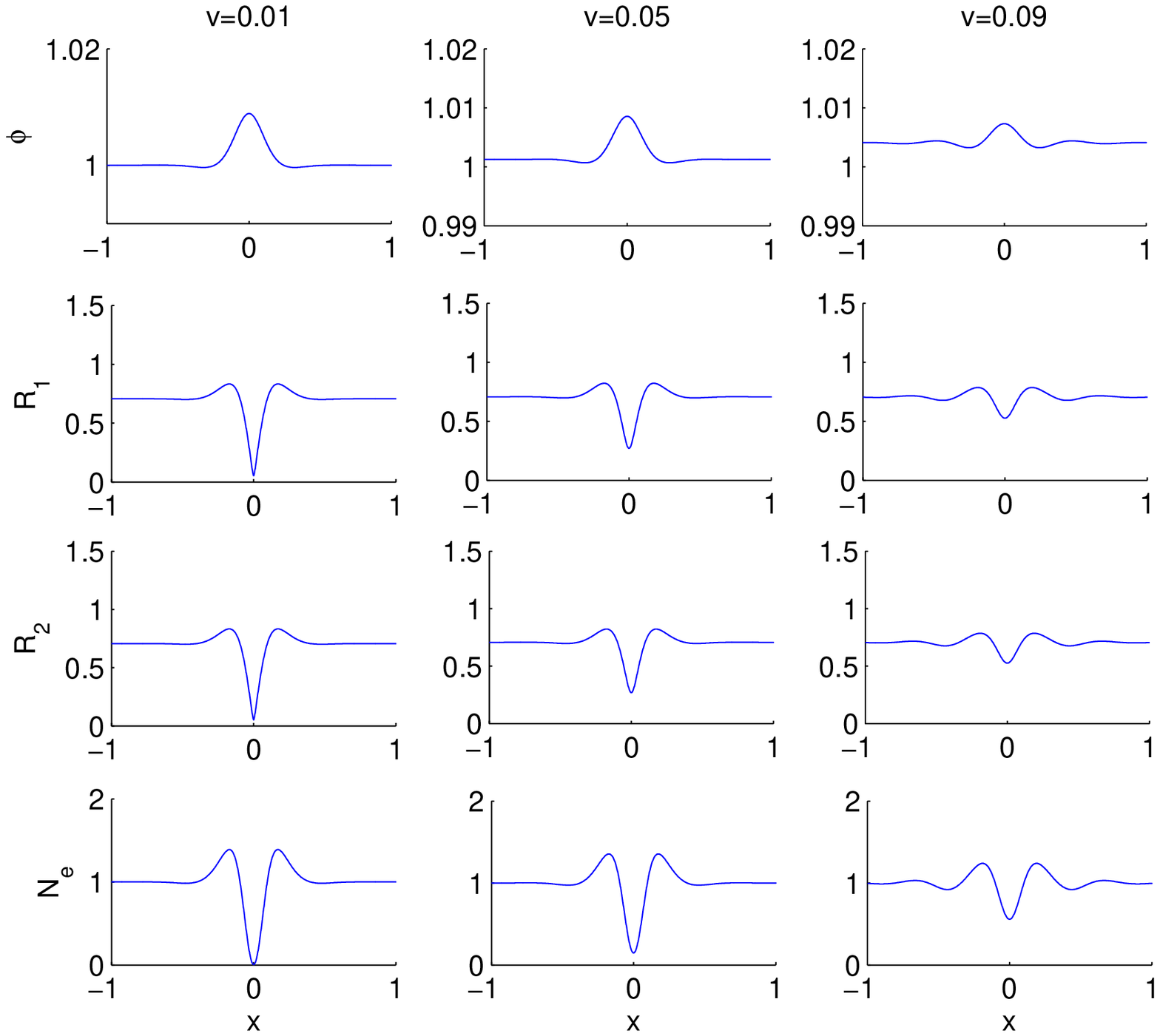}
\caption{Spatial profiles of $\phi$, $R_1$, $R_2$, and $N_e=(R_1^2+R_2^2)\phi$ (top to bottom panels), 
for $H=0.01$, and $v=0.01$ (left column), $v=0.05$ (middle column), and $v=0.09$ (right column). The solution was set to $\phi=\gamma$ and $R_1=R_2=1/\sqrt{2 \gamma}$ at the left and right boundaries. 
We see grey solitons with non-zero electron density in the center}
\label{twostream_moving}
\end{figure}

Stationary localized solutions are shown in Figs. \ref{twostream_stationary} and \ref{twostream_moving} for the non-streaming ($v=0$) and streaming ($v>0$) cases, respectively. In both cases, we found only localized solutions corresponding to $|R_1|=|R_2|=|R|/\sqrt{2}$ (where $R$ is the one-stream solution), but no other, more complicated cases. For the non-streaming solutions in Fig. \ref{twostream_stationary} we show an example with $|R_1|=|R_2|=|R|/\sqrt{2}$ in the first column (with opposite signs on $R_1$ and $R_2$). When the solution was forced to an anti-symmetric $R_1$ and a symmetric $R_2$ in space, the numerical solution converged to solutions where either $R_1$ or $R_2$ took the shape of a one-stream dark soliton, while the other part tended to zero (or as small as possible) close to the soliton. For the streaming case in Fig. \ref{twostream_moving}, we also only found localized solutions corresponding to $|R_1|=|R_2|=|R|/\sqrt{2}$. Similar as in the one-stream case, we have a maximum beam speed of $v\approx 0.1$ for the existence of localized solutions, and as the beam speed approaches this value, the amplitude of the soliton decreases and becomes oscillatory in space.

One further issue is the stability of the localized solutions in the streaming cases $v>0$. Far away from the localized solution, the plasma can be considered to be homogeneous, and one can perturb the equilibrium and study plane wave solutions proportional to $\exp(-i\Omega t+ i K x)$ for real-valued $K$ and complex-valued $\Omega$ with unstable solutions if the imaginary part of $\Omega$ is positive. In the one-stream case, studied in Section 2, all solutions were found to be stable in time, while in the two-stream case, studied in Section 4, we have $K$ for which the solutions are unstable. Hence, for the two-stream case the system is sensitive to perturbations far away from the localized structure. The general stability analysis for localized solutions can be carried out with normal mode analysis by perturbing the nonlinear equilibrium solution
of the system (\ref{r1})--(\ref{f1}) as $R_{1,2}(x,t)=R_{1,2}(x)+\widehat{R}_1(x)\exp(-i\Omega t)$, $S_{1,2}(x,t)=-\gamma m c^2 t+\sigma_{1,2}(x)+ \widehat{S}_{1,2}(x)\exp(-i\Omega t)$ and $\phi(x,t)=\phi(x)+\widehat{\phi}(x)\exp(-i\Omega t)$, and assuming that the perturbed quantities 
vanish at $|x|=\infty$. This leads to a linear eigenvalue problem  with eigenfunctions $\widehat{R}_1(x)$, $\widehat{S}_{1,2}(x)$ and $\widehat{\phi}(x)$, and eigenvalue $\Omega$. Solutions with $\Omega$ having positive imaginary parts are unstable and will grow exponentially with time.

\section{Summary and conclusion}

In this paper, we have presented a multistream model for a relativistic quantum plasma, using the Klein-Gordon model for the electrons. We have treated the one- and two-stream cases in detail. We have derived 
dispersion relations for the linear beam-plasma interactions  in the one-stream case, and for the streaming instability in the two-stream case. The system exhibits both plasma oscillations close to the plasma wave frequency that reduce to the Langmuir oscillation frequency  in the classical limit $\hbar\rightarrow 0$, 
and pair branches which do not have a classical analog. Also, there is a new instability branch for large wavenumbers of pure quantum origin. A main result of this work is the derivation of the instability condition in Eq. (\ref{insta}), which provides a natural generalization to the non-relativistic instability criterion \cite{Haas00}. Another important result is the condition (\ref{h}) for the existence of periodic (stable) oscillations and exponentially growing and decaying (unstable) steady-state oscillations, where the latter exist only below a given electron beam speed. Similar to the classical plasma case\cite{McKenzie}, this furnishes an existence condition for periodic solutions, which exist only for  cases with exponentially growing and decaying solutions. A rich variety of nonlinear solutions has been numerically found, including solitary waves with one or more electron density minima and an associated positive potential. It has been noted that the amplitude of the solitons decreased for increasing beam speeds, with increasingly oscillatory tails as the beam speed approached its maximum value for the existence of soliton solutions.

Our model can be applied to situations where the quantum statistical thermal and electron degeneracy pressure effects are small. The relative importance of these two effects can he  characterized by the degeneracy parameter $\chi = T_F/T$, where $T_F = \hbar^2 (3\pi^2 n_0)^{2/3}/(2\kappa_{B}m)$ is the Fermi electron temperature, and $\kappa_B$ is the Boltzmann constant. When $\chi < 1$,  the thermal pressure dominates, while when $\chi > 1$, the degeneracy pressure dominates. Our model is applicable when $\kappa_B T \ll mv^2$ for $\chi > 1$ and $\kappa_B T_F \ll mv^2$ for $\chi < 1$.  Strong coupling (collisional) effects can be neglected when the coupling constant $\Gamma = e^2 n_{0}^{1/3}/(4\pi\varepsilon_{0}\kappa_{B}T)$ (for $\chi < 1$) or $\Gamma = e^2 n_{0}^{1/3}/(4\pi\varepsilon_{0}\kappa_{B}T_F)$ (for $\chi > 1$) is small. 
For $\chi > 1$, the Pauli blocking further helps to reduce the effect of collisions \cite{Ashcroft76}.
The pair creation phenomena have been neglected here, because our model excludes quantized fields.
Hence, $\hbar\omega_p$ needs to be much smaller than $2mc^2$ \cite{Tsytovich61}.
Working out the weak coupling and no quantized field assumptions, formulated as $\Gamma < 1$ and $\hbar\omega_{p}/(2mc^2) < 1$, we have  (using SI units)

\begin{equation}
\quad \log_{10}T > \frac{1}{3}\log_{10}n_0 - 4.8,  \quad \mbox{if} \quad \chi < 1,
\end{equation}

\begin{equation}
\log_{10}n_0 > 28.8, \quad \mbox{if} \quad \chi > 1,
\end{equation}
and

\begin{equation}
  \log_{10}n_0 < 38.9.
\end{equation}
as the condition for the applicability of our theoretical model.

Finally, it should be noted that our investigation neglects electron-one-half spin effects, which are justified since we used an unmagnetized quantum plasma model, and hence there are no spin couplings to a magnetic field. In conclusion, we stress that the present investigation of linear and nonlinear effects dealing with relativistic electron beams in a quantum plasma is relevant for high intensity laser-plasma interaction experiments \cite{Mourou}, white dwarf stars \cite{Galloway,Winget}, and neutron stars \cite{Anderson03,Anderson04,Samuelson10}, where both quantum and relativistic effects could be 
important.

\begin{acknowledgments}
This work was supported by Conselho Nacional de Desenvolvimento Cient\'{\i}fico e Tecnol\'ogico (CNPq),
as well as by the Deutsche Forschungsgemeinschaft through the project SH21/3-2 of the Research Unit 1048.
\end{acknowledgments}

\end{document}